\documentclass[prb,twocolumn,showpacs,floatfix]{revtex4}%
\usepackage{graphicx}
\usepackage{bm}
\begin{document}

\title{Electronic transport in quasi-one-dimensional arrays of gold nanocrystals}

\author{Klara Elteto$^1$, Xiao-Min Lin$^2$, and Heinrich M. Jaeger$^1$}
\affiliation{$^1$James Franck Institute and Department of Physics,
   University of Chicago, Chicago, IL 60637\\
$^2$Materials Science Division, Chemistry Division and Center for
Nanoscale Materials, Argonne National Laboratory, Argonne, IL
60439}

\date{\today}

\begin{abstract}
We report on the fabrication and current-voltage ({\em IV})
characteristics of very narrow, strip-like arrays of metal
nanoparticles.  The arrays were formed from gold nanocrystals
self-assembled between in-plane electrodes.  Local cross-linking
of the ligands by exposure to a focused electron beam and
subsequent removal of the unexposed regions produced arrays as
narrow as four particles wide and sixty particles long, with high
degree of structural ordering. Remarkably, even for such
quasi-one-dimensional strips, we find nonlinear, power-law {\em
IV} characteristics similar to that of much wider two-dimensional
(2D) arrays. However, in contrast to the robust behavior of 2D
arrays, the shape of the {\em IV} characteristics is much more
sensitive to temperature changes and temperature cycling.
Furthermore, at low temperatures we observe pronounced two-level
current fluctuations, indicative of discrete rearrangements in the
current paths. We associate this behavior with the inherent high
sensitivity of single electron tunneling to the polarization
caused by the quenched offset charges in the underlying substrate.
\end {abstract}

\pacs{73.23.-b, 73.22.-f, 73.23.Hk}

\maketitle

\section{Introduction}

Metal or semiconductor nanoparticles and complex structures built
from them through self-assembly \cite{whitesides,redl,lopes,liao}
provide model systems for further investigation of physics on the
mesoscopic scale, where quantum confinement and Coulomb charging
set the relevant energy scales. \cite{imry} Electronic properties
of single particles \cite{vonDelft} as well as two- and
three-dimensional superlattices of nanocrystals
\cite{gs,bawendi,parthasarathyIV} have been extensively studied
over the last decade. Here we address what happens to electronic
transport as the one-dimensional limit is approached.

Electronic transport through an array of small metal particles
separated by nanoscale gaps is determined by the interplay between
single electron charging of an individual particle and tunneling
between adjacent particles.  In the presence of charge disorder
due to quenched impurities in the insulating substrate under the
array, \cite{middleton} this interplay leads to highly non-Ohmic
current-voltage {\em IV} characteristics. Theory and simulations
predict that, at sufficiently low temperatures, no current flows
below a voltage threshold, $V_t$, while above it the current
follows a power-law, $I \sim (V-V_t)^{\zeta}$ , with $\zeta = 1$
in 1D and between 5/3 and 2 in 2D.
\cite{middleton,reichhardt,rendell}

Experimentally, true 1D chains are difficult to achieve and, so
far, results have been obtained only on quasi-1D structures with
significant amount of disorder in the particle arrangements.
Narrow chains of carbon nanoparticles \cite{bezryadin} showed
sample-dependent scaling exponents between 1 and 2.3.
Electron-beam written 100nm wide multilayers of $\rm{Au_{55}}$
particles \cite{clarke} exhibited $\zeta = 1.6$, and experiments
on strips of 1.8nm gold clusters \cite{shirley} gave $\zeta = 3$.
This spread in exponents is in sharp contrast with the situation
for well-ordered, 2D Au monolayer nanocrystal arrays. Experiments
on highly ordered 2D nanocrystal arrays found $\zeta =
2.25\pm0.1$, a value that is temperature-independent and highly
reproducible from sample to sample.
\cite{parthasarathyIV,parthasarathyTDep}

These results highlight two key issues.  First, to what extent is
the spread in the measured scaling exponents caused by the charge
disorder in (quasi-) 1D structures rather than to structural
disorder and fabrication details; second, how narrow does an array
have to be in order for its current-voltage characteristics to
cross over from a nonlinear, 2D behavior to the a linear response
with $\zeta = 1$?  To address these issues, we performed
systematic measurements on quasi-1D strips fabricated from
well-ordered monolayer superlattices. These structures also allow
us to observe explicitly the effects of quenched charges in the
substrate through measuring the current fluctuation at fixed
temperature and the changes in the current-voltage characteristics
after temperature-cycling the array.

\section{Samples}

Gold nanocrystals of 5.5nm diameter with 5-7\% dispersity were
synthesized as described in Ref. \onlinecite{linSynthesis} through
a digestive ripening method and suspended in toluene. A droplet of
this colloidal suspension was then deposited onto
$\rm{Si_3N_4}$-coated Si substrates with prefabricated, ~15nm
thick in-plane chromium electrodes ($\rm{2 \mu m}$ wide with 500nm
gap, i.e., about 60 particles across). By controlling the particle
concentration, solvent evaporation rate and concentration of
excess alkanethiol ligands in the solution, this drop drying
procedure was used to form highly ordered nanocrystal monolayers.
\cite{linOrder} Alternatively, a water droplet was first deposited
onto the substrate and the colloidal suspension was allowed to
creep up at the water-air interface, forming a thin layer which
then draped itself over the substrate as the water slowly
evaporated. \cite{eah}  Both methods allow nanocrystals to
self-assemble into a compact 2D monolayer at the liquid-air
interface before the solvent dewetting occurs.

After solvent evaporation, the quasi-1D array was fabricated by
exposing the corresponding monolayer regions to a finely focused
electron beam in a scanning electron microscope (SEM).
\cite{werts,lohau,linEBW} Line doses of $\rm {8-25nC/cm}$ or area
doses of $\rm{10mC/cm^2}$ were used to e-beam-write lines or
rectangles at 30kV. These exposure parameters effectively
cross-link the interstitial ligands but do not affect the
nanoparticles' shape and relative position. \cite{werts} The
exposed regions become resilient to heated toluene, which can be
used to wash away unexposed particles. Overall, this method works
like a positive e-beam resist in that the final structures
correspond to the exposed areas.  Back-etched $\rm{Si_3N_4}$window
areas in the substrate allowed for inspection by transmission
electron microscopy (TEM). In our experiments, all the samples
were inspected by TEM after completing the electronic
measurements.

\begin{figure}
\begin{center}
\includegraphics[width=6.8cm]{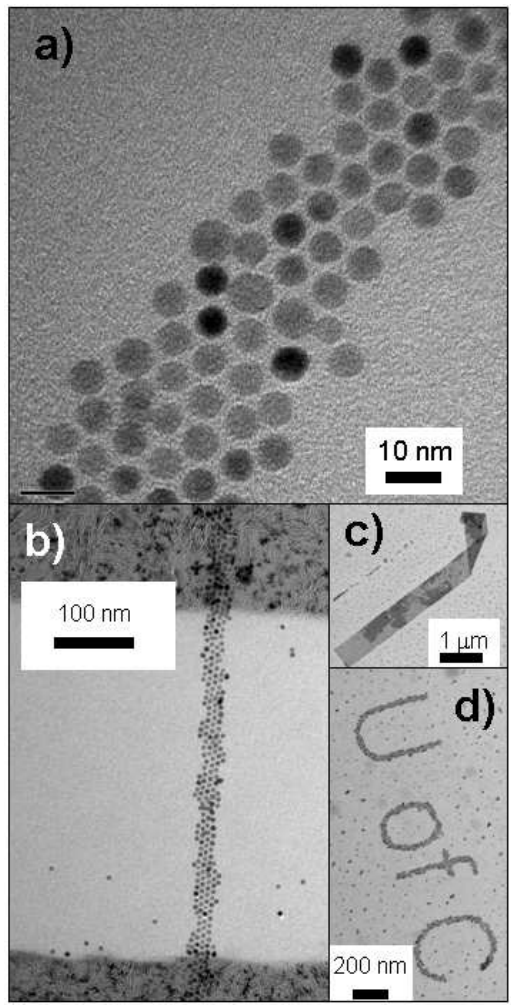}
\end{center}
\caption{Transmission electron micrographs of structures patterned
from an extended nanoparticle monolayer using electron beam
writing. a) Large magnification image of a narrow strip, showing
the integrity of the particle arrangement after e-beam exposure.
b) 30nm wide, quasi-1D array between planar Cr electrodes. Gaps
between the particles are 1.3nm on average. c) Beam-written array
formed using a colloid containing excess dodecanethiol ligand. The
array folds over, as shown here, rather than break apart during
aggressive agitation in heated toluene, indicating e-beam-induced
cross-linking of the ligands. d) Arbitrary patterns with both
straight and round sections are possible.}
\label{ftem}
\end{figure}

Figure \ref{ftem} shows TEM images of structures fabricated by
this method. The shape of the final patterns depends only on the
ability to control the position and movement of the electron beam,
while the ultimate resolution appears to be limited primarily by
the e-beam spot size and the degree of ordering in the original
monolayer. Panel a gives the details of a narrow strip fabricated
through this method. Panel b shows one of the measured arrays with
the Cr electrodes visible. All samples were structurally ordered
in the sense that they retained the close-packed particle
arrangement of the monolayer, showed few defects, exhibited a
uniform inter-particle spacing, and had well-defined edges with a
roughness of 1-2 particles (in many cases unavoidable because of
the mismatch between the particle lattice orientation and the
direction of e-beam writing).

Six arrays were beam-written as single lines and had average
widths of 30nm, or four particles across. Two arrays were
patterned as slightly wider rectangles, on average seven particles
across. Isolated, localized patches where particles formed a
second layer made up less than 15\% of the area in each of the
eight arrays.

All samples were mounted in a shielded cryostat and connected with
low-noise coaxial cables to a Keithley 6340 sourcemeter.
Temperature was controlled from 8K to 120K, with stability better
than 0.1\%. We measured the current while ramping the bias voltage
between -20V and +20V at rates of 50mV/sec or less.  The rms
current noise of the set-up was about 10fA.

As a control study, we measured full 2D monolayers before and
after e-beam writing and washing. We found that the voltage
threshold at a given temperature, and the scaling exponent did not
change by more than the fitting uncertainty. This shows that the
alkanethiol ligands act as mechanical spacers and, cross-linked or
not, do not otherwise influence the properties of the tunneling
barrier between the neighboring particles.

\section{Results and Discussion}

\begin{figure}
\begin{center}
\includegraphics[width=8.6cm]{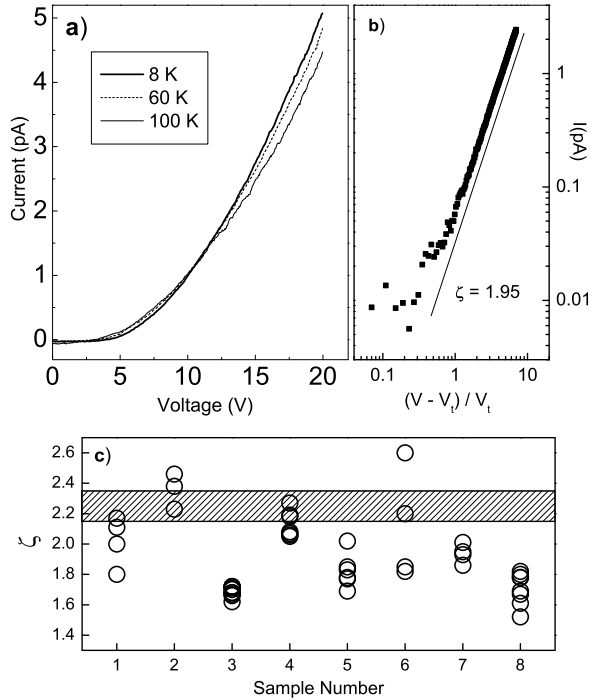}
\end{center}
\caption{Current-voltage ({\em IV}) characteristics of a
4-particle-wide array. a) The shape of the {\em IV} curves changes
with temperature, unlike what is observed for wider, 2D arrays. b)
Log-log plot of the {\em IV} curve for the same array as in (a) at
20K with best-fit voltage threshold $V_t(\rm{20K) = 2.5V}$. The
line corresponds to a scaling exponent $\zeta = 1.95$. c)
Temperature-dependent values of $\zeta$ for all eight samples
measured. The hatched area gives the standard deviation in the
observed values for $\zeta$ in 2D arrays.} \label{fIV}
\end{figure}

Typical current-voltage curves at different temperatures are shown
in Fig. \ref{fIV}a. Data from all measured arrays follows
power-law scaling, $I \sim (V-V_t)^{\zeta}$, with voltage
thresholds, $V_t$, in the several-volt regime and scaling
exponents, $\zeta$, around two. Figure \ref{fIV}b shows the extent
of the power-law scaling. For each temperature, we determined the
voltage threshold and the scaling exponent from best straight-line
fits on a log-log plot. On closer inspection, Fig. \ref{fIV}a also
shows that the shape of the {\em IV} traces changes with
temperature. This is in contrast to what has been observed for
wider, 2D arrays, where the scaling exponent is
temperature-independent and the {\em IV} curves can be collapsed
simply by a translation along the voltage axis.
\cite{parthasarathyTDep} In Fig. \ref{fIV}c, we plot both the
extent of the temperature-induced variation in $\zeta$ for a given
sample and the variation of $\zeta$ for different samples. For
comparison, we also show the very small variation of the scaling
exponent obtained from a series of 2D arrays fabricated with the
same Au nanoparticles. \cite{parthasarathyIV}

For $V_t$, on the other hand, we find behavior similar to what has
been previously observed in 2D arrays
\cite{parthasarathyTDep,boos} and in 1D chains, \cite{bezryadin}
namely an approximately linear decrease with increasing
temperature. At zero temperature, the voltage threshold can be
expressed as $V_t(0) = \alpha NV_0$, where $N$ is the number of
particles along the length of the array and $eV_0$ corresponds to
the typical electrostatic energy cost associated with single
electron tunneling between neighboring particles in the array.
\cite{parthasarathyTDep} Because quenched charge disorder leads to
a distribution of local energy costs, tunneling will occur along
those current paths that minimize penalties and circumvent sites
with large local charging energies.  The extent to which this
happens is measured by the prefactor $\alpha$ which depends on the
array geometry and dimensionality.

For 2D close-packed arrays, characterized by a large coordination
number and thus a multitude of possible detours, theory predicts
$\alpha = 0.226$, while $\alpha = 1/2$ for 1D chains.
\cite{middleton,parthasarathyTDep} To find $\alpha$
experimentally, we obtained $N$ directly from TEM images and
established $V_t(0)$ by extrapolating the temperature-dependent
threshold to zero temperature.  Using the procedure described in
Ref. \onlinecite{elteto}, $V_0$ was calculated from the average
radius and center-to-center spacing of the particles in each
array, as determined by TEM. Our quasi-1D strips exhibit values
between the 1D and 2D theoretical prediction, giving $\alpha =
0.34 \pm 0.09$ for the eight samples measured.

The observed values for $\alpha$ and $\zeta$ demonstrate that,
even at average widths of only four particles, transport along the
arrays has not yet reached the 1D limit.  The nonlinear shape of
{\em IV} curves shows that the current paths are still able to
meander and branch significantly, which is the characteristic
behavior of 2D transport.

Nevertheless, quasi-1D arrays allow for only a very limited number
of parallel paths. We therefore expect that these limited number
of current-carrying paths will be inherently more susceptible to
changes in the configuration of quenched offset charges in the
underlying substrate.

\begin{figure}
\begin{center}
\includegraphics[width=7.7cm]{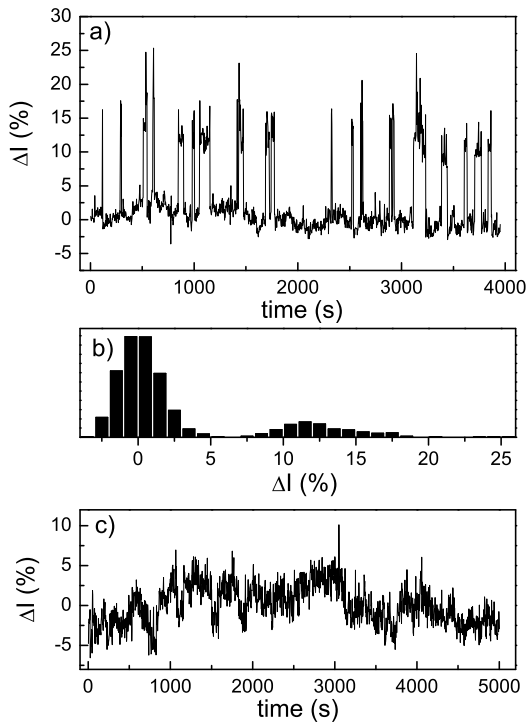}
\end{center}
\caption{Current fluctuations at fixed bias voltage. a) Percentage
change in current as a function of time for an array at 8K and
-12V bias. At the large jumps, the current value changes by as
much as 1pA between successive data-points (1 second). b)
Histogram of the data in (a). c) Percentage change in current as a
function of time for the same array at 100K and -12V bias.}
\label{ffluct}
\end{figure}

This is borne out by measurements of the fluctuations in the
transport current at low temperatures under a fixed bias voltage.
We observe clear evidence for random switching between a few
well-defined states of the system.  The $I(t)$ trace at 8K in Fig.
\ref{ffluct}a shows that the current fluctuates on the order of
several minutes between two steady states, which also are easily
identified from the histogram in Fig. \ref{ffluct}b. The magnitude
of the current change in this figure, about 12\%, corresponds to
one out of four parallel conduction channels in the array changing
its transmission by 50\%.

As the temperature increases between 8K and 60K, the frequency of
this telegraph noise increases and, above 60K, distinct current
states can no longer be observed (Fig. \ref{ffluct}c).
Qualitatively, this change in behavior is expected if the number
of active conduction channels in the array depends on the
configuration of quenched offset charges in the substrate. With
increasing thermal energy, these offset charges can switch more
rapidly between different trapping states, inducing the opening
and closing of conduction paths. Similar stochastic switching of
current was also observed in much wider 2D arrays (with width of
$\rm{2 \mu m}$). However, these 2D arrays did not exhibit bimodal
switching even at 8K.

\begin{figure}
\begin{center}
\includegraphics[width=8.3cm]{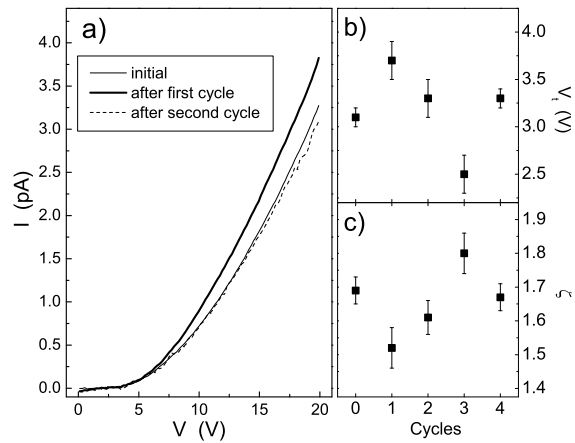}
\end{center}
\caption{The effect of temperature cycling.  a) Current as a
function of voltage for a 4-particle-wide array at 20K after
several temperature cycles to 200K. b) Voltage threshold, $V_t$,
of the array obtained from power-law fitting of {\em IV} curves.
c) Scaling exponent, $\zeta$ , of the {\em IV} curves. The error
bars indicate the fitting uncertainty. } \label{fTcycle}
\end{figure}

Further evidence that rearrangements of trapped offset charges
determine the behavior of the mobile electrons comes from
temperature-cycling experiments. In these experiments, we measure
the current at 20K, warm up the array to 200K for at least an hour
and, and upon cooling back to 20K, measure the current again (Fig.
\ref{fTcycle}a).  Such temperature cycles are done several times
for the same sample. We find that, in response, the threshold
voltage and scaling exponent change randomly (Fig.
\ref{fTcycle}b,c), varying significantly more than the fitting
uncertainty given by the error bars. In fact, the changes induced
by temperature cycling are of similar magnitude as the
sample-to-sample variations (Fig. \ref{fIV}c). This variability
contrasts with the robust behavior we find in 2D arrays that are a
few hundred particles wide. For 2D arrays, even after temperature
cycling, $\zeta$ stays within the range indicated in Fig.
\ref{fIV}c.

\section{Conclusions}

In summary, our results demonstrate the important role played by
quenched or trapped offset charges in determining the arrays'
overall transport behavior.  Being able to control the structural
integrity of the arrays down to widths of 4 particles, we conclude
that the large variability in threshold voltages and scaling
exponents seen in our experiments as well as by others
\cite{bezryadin,clarke,shirley} is inherent to the quasi-1D nature
of the samples. The fact that there are fewer conduction paths in
narrow strips as compared to the extended 2D arrays implies a high
sensitivity to the change of the quenched charge disorder
``landscape''. This charge reconfiguration is time dependent and
induced by thermal activation.

In a 2D array by contrast, changes in the charge configuration and
tunneling paths average out, giving a time- and
temperature-independent scaling exponent that varies little from
array to array.  Thus, while quenched charge disorder gives rise
to the nonlinear current-voltage characteristics in the first
place, quasi-1D structures at any given time sample only a limited
range of disorder configurations.  This has a significant
implication for device applications in which a high degree of
reproducibility from sample to sample is desired.  In those cases,
it will be necessary to use wider nanocrystal arrays in which
there is sufficient spatial averaging over different trapped
charge disorder configurations.

\begin{acknowledgments}
The authors thank Terry Bigioni and Thu Tran for illuminating
discussions, Sang-Kee Eah for sharing the water-drop-depositing
technique, and Qiti Guo and Robert Josephs for technical help with
the electron microscopes.  This work was supported by the UC-ANL
Consortium for Nanoscience Research, by the NSF MRSEC program
under DMR-0213745, and by the DOE, Basic Energy Sciences-Materials
Sciences, under contract \#W-31-109-ENG-38.
\end{acknowledgments}

\end {document}